\documentclass[twocolumn]{aa} 
\usepackage{graphicx}
\usepackage{txfonts}
\usepackage{natbib}
\bibpunct{(}{)}{;}{a}{}{,} 

\begin{document}
%
\title{Determining global parameters of the oscillations of solar-like stars}
 \titlerunning{Determining global parameters of the oscillations of solar-like stars}

   \author{S. Mathur\inst{1,2} \and 
                R.~A. Garc\'\i a\inst{1,3} \and
   		C. R\'egulo\inst{4,5} \and
		O.~L. Creevey\inst{4,5}\and
		J. Ballot\inst{6} \and
		D. Salabert\inst{4,5} \and
		T. Arentoft\inst{7} \and
		P.-O.~ Quirion\inst{7} \and
		W.~J. Chaplin\inst{8} \and
		H. Kjeldsen\inst{7}
	}
   \offprints{savita.mathur@cea.fr}
   \institute{Laboratoire AIM, CEA/DSM -- CNRS - Universit\'e Paris Diderot -- IRFU/SAp, 91191 Gif-sur-Yvette Cedex, France
   \and Indian Institute of Astrophysics, Koramangala, Bangalore 560034, India
   \and GEPI, Observatoire de Paris, CNRS, Universit\'e Paris Diderot; 5 place Jules Janssen, 92190 Meudon, France
   \and Universidad de La Laguna, Dpto de Astrof\'isica, 38206, Tenerife, Spain
   \and Instituto de Astrof\'\i sica de Canarias, 38205, La Laguna, Tenerife, Spain
   \and Laboratoire d'Astrophysique de Toulouse-Tarbes, Universit\'e de Toulouse, CNRS, F-31400, Toulouse, France
  \and Danish AsteroSeismology Centre (DASC), Department of Physics and Astronomy, Aarhus University, 8000 Aarhus C, Denmark
   \and School of Physics and Astronomy, University of Birmingham, Edgbaston, Birmingham B15 2TT, UK 
   }

   \date{Received 2009; accepted }

 \abstract
    {Helioseismology has enabled us to better understand the solar interior, while also allowing us to better constrain solar models. But now is a tremendous epoch for asteroseismology as space missions dedicated to studying stellar oscillations have been launched within the last years (MOST and CoRoT). CoRoT has already proved valuable results for many types of stars, while Kepler, which was launched in March 2009, will provide us with a huge number of seismic data very soon. This is an opportunity to better constrain stellar models and to finally understand stellar structure and evolution.}
   {The goal of this research work is to estimate the global parameters of any solar-like oscillating target in an automatic manner. We want to determine the global parameters of the acoustic modes (large separation, range of excited pressure modes, maximum amplitude, and its corresponding frequency), retrieve the surface rotation period of the star and use these results to estimate the global parameters of the star (radius and mass).}
   {To prepare for the arrival and the analysis of hundreds of solar-like oscillating stars, we have developed a robust and automatic pipeline, which was partially adapted from helioseismic methods. The pipeline consists of data analysis techniques, such as Fast Fourier Transform, wavelets, autocorrelation, as well as the application of minimisation algorithms for stellar-modelling.}
   {We apply our pipeline to some simulated lightcurves from the asteroFLAG team and the Aarhus-asteroFLAG simulator, and obtain results that are consistent with the input data to the simulations. Our strategy gives correct results for stars with magnitudes below 11 with only a few 10\% of bad determinations among the reliable results. We then  apply the pipeline to the Sun and three CoRoT targets.
In particular we determine the large separation and radius of the Sun, HD49933, HD181906, and HD181420.}
  {}


   \maketitle
   
\section{Introduction}

During the past 30 years, helioseismology has proven its ability to study the structure and dynamics of the solar interior in a stratified way. These seismic tools allow us to infer some physical quantities as a function of the radius:  the sound-speed \citep[e.g.][]{BasJCD1997}, the density \citep[e.g.][]{STCCou2001}, the rotation profile in the convective zone \citep[e.g.][]{ThoToo1996}, or in the radiative zone \citep[e.g.][]{2008SoPh..251..119G} are some well known examples. But helioseismology has also provided other quantities such as the position of the base of the convection zone \citep{JCDGou1985, BalSTC2004}, and the photospheric Helium abundances \citep[][]{1991Natur.349...49V,1995MNRAS.276.1402B}. These observational constraints improved the standard solar models significantly. For instance, with the solar sound speed \citep[e.~g.][and references therein]{2007ApJ...668..594M} the agreement between theory and seismic observations is better than a few parts per hundred \citep[using the recent revision of solar abundances from][]{2005ASPC..336...25A} or even a few parts per thousand \citep[with the old solar abundances of][]{1993A&A...271..587G}. 

The Sun is a fundamental calibrator of stellar evolution \citep[e.g.][]{2009IAUS..258..431C}, but observations of many other stars -- covering the HR diagram through asteroseismology -- will allow testing stellar evolution and dynamo theories under many different conditions \citep{2008A&A...485..813C}. However, compared to the Sun, less information will be available from asteroseismology as only low-degree modes (those with a small number of nodal lines at the surface of the star) will be accessible due to the absence of spatial resolution in the observations. This reduces the accuracy and spatial resolution of the inversions. On the other hand, some stars offer the possibility of observing gravity modes, thus giving access to the structure and dynamics of their radiative interiors \citep[e.~g.][]{2008A&A...484..517M}. 

Nowadays, we are approaching a crucial period of asteroseismic observations. Several campaigns of ground-based observations have been organised \citep[e.~g.][]{2003PASA...20..203B,2008ApJ...687.1180A}. Besides, several space missions have already been observing   -- for more than 3 years -- solar-like oscillating stars (whose acoustic oscillations are excited stochastically by convection): the American satellite WIRE \citep[Wide-Field Infrared Explorer,][]{BruKje2005}, the Canadian satellite MOST \citep[Microvariability and Oscillations of STars,][]{2003PASP..115.1023W}, and the French-European-Brazilian mission CoRoT \citep[Convection, Rotation and planetary Transits, see e.g.][]{2008Sci...322..558M}. With the nominal observation cycle of CoRoT (changing the pointing field every six months), only a very few solar-like oscillating stars are observed each year \citep[see for example:][]{2008A&A...488..705A,2009A&A...506...41G,2009A&A...506...51B}, allowing us to analyse each individual target very carefully. The launch of Kepler \citep{2009IAUS..253..289B} on March 7, 2009, heralds a new era for asteroseismology, in that it will increase the number of solar-like stars observed for asteroseismology  by two orders of magnitude. Kepler, which will search for habitable exoplanets, also has a wide asteroseismic programme, the Kepler Asteroseismology Investigation \citep[KAI,][]{2008JPhCS.118a2039C}. Because it will observe a large number of stars for asteroseismology, Kepler presents many challenges for the efficient analysis of the data. The nominal Kepler lifetime is expected to be 3.5 years allowing very accurate asteroseismic  studies of the same field all or the entire mission and the possibility of measuring stellar cycles   
\citep{2007MNRAS.379L..16M,2009arXiv0906.5441K}.


Concerning the asteroseismic part, a survey phase will take place over the first 10 months of Kepler nominal operations \citep[e.~g.][]{ChapApp10} The main goal of this phase is to select the best stars that will be targeted during the rest of the nominal mission for, at least, two and a half years. Each star will be observed for one month at a time, putting constraints on the way how to extract the seismic parameters. For brighter targets we will be able to get individual frequencies, but for the majority of stars (fainter ones) this will not be the case and one must rely on extracting mean global parameters
\citep[see for example the analysis of a 26 days of CoRoT observations on a solar-like target by][]{2009A&A...506...33M}.  

About a thousand stars are going to be observed during the survey phase at a one-minute cadence. Consequently, people working in the Kepler Asteroseismic Science Consortium (KASC) will have to analyse several hundreds of stars every month. This is the reason why we need to change our single-star analysis approach to an automatized pipeline that will provide global-seismic parameters as well as other first estimations of structural parameters like their radius or mass. 
Some teams have developed their own pipeline \citep{2009arXiv0911.2612H,2009arXiv0910.2764H} within the AsteroFLAG group to be ready to analyse the Kepler data.

The objective of this paper is to describe the pipeline we have developed to achieve this goal. The global parameters of the oscillations that are some of the output of our pipeline can be used by modelers to constrain their stellar models \citep[e.~g.][]{2009ApJ...699..373M,2009ApJ...700.1589S}

Sect.~2 describes the goals and methods of the different packages of our pipeline dedicated to the determination of the p-mode global parameters. Then in Sect.~3, we explain how we retrieve the radius of the stars from these parameters. To illustrate that our pipeline works properly as well as its limits, we applied it to some simulated data by asteroFLAG team \citep{2008AN....329..549C} and by a combined Aarhus/asteroFLAG simulator, to the Sun, and to three of the CoRoT targets in Sect.~4. Finally, we discuss the results of our pipeline and the strategy adopted.

\section{Pipeline package description}

\begin{figure}[htb*]
\begin{center}
\includegraphics[width=9cm,trim=4mm 15mm 3mm 40mm]{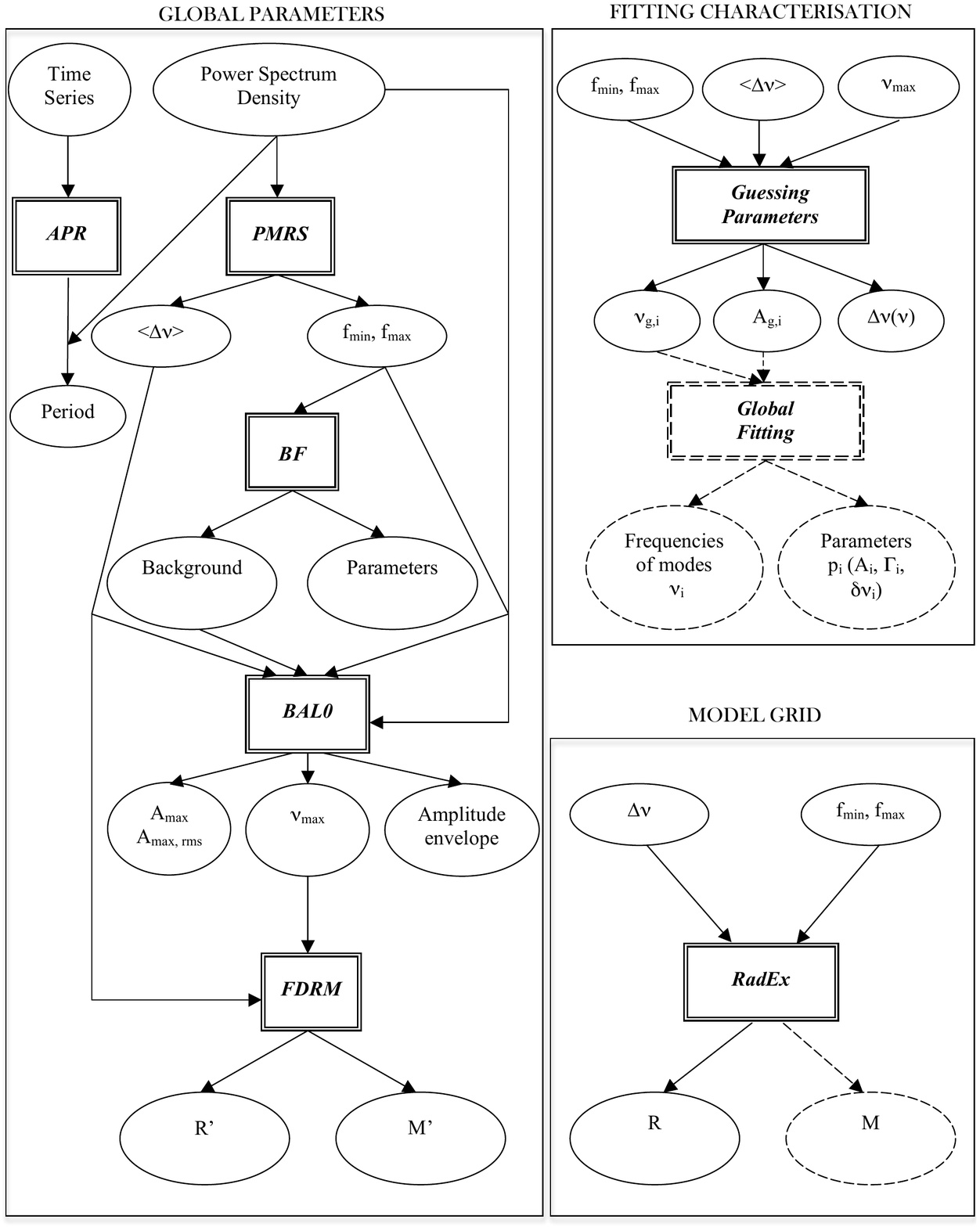}
\caption{Workflow diagram of the \emph{A2Z} pipeline to extract the global parameters of stellar oscillations and the  characteristics of the modes and to determine of the mass and radius through the comparison with a pre-calculated grid of models. The rectangles represent the code packages and the input/output are in the circles (each output has its corresponding error bar). The dashed lines correspond to the packages that are not described in this paper.}
\label{workflow}
\end{center}
\end{figure}

The automatic pipeline that we have developed is called \emph{A2Z}. It is divided into three main parts (see Fig.~\ref{workflow} for a detailed view of the workflow diagram). The first one is dedicated to the study of the global parameters of solar-like oscillating stars finishing by computing a first estimation of the mass and the radius of the star deduced from empirical scaling laws.
The second part corresponds to the extraction of p-mode properties through the fitting of individual modes. 
The third part uses a pre-calculated grid of evolutionary models where we give as input previously determined parameters such as the large separation and the frequency range of the p-mode excess to obtain a better estimation of the radius and mass.

The starting point of our \emph{A2Z} pipeline is either the time series or the power spectrum density (PSD) 
expressed in ppm$^2/\mu$Hz. A Lomb-Scargle algorithm is used depending on the regularity of the sampling rate of the time series.

The package APR (Sect.~2.1, average photospheric rotation) gives an estimation of the surface rotation period of the star analysing either the low-frequency part of the PSD or looking for the time-frequency signature of this periodicity using wavelets.
 
  PMRS (Sect.~2.2, p-mode range search), uses two different methods to estimate the mean large separation ($\langle \Delta \nu \rangle$) and gives the range of the p-mode excess power ($f_{\rm min}$ and $f_{\rm max}$). These latter values are then used to fit the background in the PSD with the package BF (Sect.~2.3, background fitting). Then, the third package, BAL0 (Sect.~2.4, bolometric amplitude per $\ell$=0), calculates the amplitude envelope and gives the maximum bolometric amplitude per radial mode ($A_{\rm max}$) as well as the corresponding frequency ($\nu_{\rm max}$). Finally, with these asteroseismic parameters, we can estimate the radius (R') and the mass (M') of the stars using empirical scaling laws with the package FDRM (Sect.~2.5, First Determination of the Radius and the Mass). 
  
 For the package dedicated to the fitting of the modes, we need a table with the guesses of the mode frequencies. The package Guessing Parameters (Sect.~2.6) uses the global parameters of the modes obtained previously to calculate a table of the guesses for the frequencies ($\nu_{g,i}$) and the amplitudes ($A_{g,i}$) (where $g$ is used for the guesses and $i$ is the mode index) as well as a first estimation of the variation of the large separation with frequency ($\Delta \nu$($\nu$)).
 
 Another package RaDex (Sect.~3, radius extractor) based on a grid of stellar models also estimates the radius of the stars using $\langle \Delta \nu \rangle$, $f_{\rm min}$, and $f_{\rm max}$. All the outputs of the different boxes are calculated with their uncertainties, $\epsilon$.

 If the signal-to-noise ratio (SNR) is high enough, the global parameters and the guesses are used by the maximum-likelihood Global Fitting code to infer the characteristics of each mode: the central frequency $\nu_i$, the amplitude A$_i$, the width $\Gamma_i$, and the rotational splittings $\delta \nu_i$. This global fitting code will be used only on a few stars during the survey phase of Kepler. However, the characterisation of individual p modes is out of the scope of this paper. It has already been tested on the Sun a large number of times and applied to several CoRoT targets \citep[e.~g.][]{2008A&A...488..705A,2009A&A...506...51B, 2009A&A...506...41G}.

In the following sections, we describe the different packages and apply them to solar-like stars simulated under typical asteroseismic conditions. In some packages, we use different methods to cross check the results.

\subsection{Average Photospheric Rotation}

One of the first things that we want to study is the surface rotation period of the star. Thanks to seismology, we can independently determined the inclination 
from the relative amplitude of mode components inside multiplets 
\citep{2003ApJ...589.1009G}. Unfortunately, the seismic estimates of the 
rotational splitting and of the inclination $i$ --simultaneously 
obtained by fitting observed oscillation spectra-- are strongly 
correlated for many of solar-like stars due to their short mode lifetime 
\citep[for more details and discussion, see][]{BalGar2006,2008A&A...486..867B}. Thus, an independent determination of one of these two parameters (rotation or inclination angle) could help during the fitting procedure of the p-mode parameters to establish, for example, some a priori constraints \citep[e.g.][]{2008CoAst.157...98B,2009A&A...507L..13B}. However, it is important to remember that the rotational splitting of the mode is an average value of rotation inside the cavity in which the mode propagates and can be different from the surface rotation if there is strong differential rotation in the interior of the star.

To infer the rotation period of the stellar surface several methods can be used: studying the PSD or using the wavelets.
The classical method consists in analysing the PSD at very low frequency. We look at the highest peaks below 5~$\mu$Hz (a value that can be changed to take into account more rapid rotators). However, with this method it is difficult to distinguish between the fundamental peak of the stellar surface rotation period and its first harmonics. 

The other method that we use in this work consists of applying a time-frequency analysis using the wavelets \citep{1998BAMS...79...61T}. Before applying the wavelet tools, we reduce the number of points by rebinning the data  by a boxcar function  to speed up the calculation. We also filter the data to have only the low-frequency region. Then we use the Morlet wavelet, which is the product of a Gaussian and a sinusoid, thus a finite pattern and change the scale of the wavelet, i.e. its frequency (or period). By sliding it along the time series we obtain the wavelet power spectrum (WPS), which is the result of the correlation between the wavelet and the data. This method has the advantage of lifting the uncertainty on the rotation period (see Sect.~4.2). This tool has also been extensively used to study solar and stellar activity \citep{2008arXiv0810.1803M}.
 Besides, a very conservative approach has been followed by defining the cone of influence (COI) to determine the limit of a reliable result: a periodicity should be seen at least four times to have more confidence on the value.

An example of the WPS is plotted on Fig.~\ref{fig5b} using solar data. We calculate the sum of the WPS along time for each period (or frequency) of the wavelet. The result, which is the global wavelet power spectrum (GWPS), is represented as a function of the period (or frequency of the wavelet) (see Fig.\ref{fig5b}, right panel). We overplotted as well the threshold for a confidence level of 95\% (chosen for our study). Any peak that is above this threshold has more than 95\% probability not to be due to noise. The surface rotation period of the star is easily identified with this confidence level. 

\subsection{P-Mode Range Search and large separation estimation}

We study the stellar oscillations spectrum by searching for the frequency range where the p modes present most of their power in the PSD. This search begins with the estimation of the mean large separation, $\langle \Delta \nu \rangle$. 

It is important to have two independent methods to obtain the spacing, to cross-check the results, and to be able to have reliable results. 

\subsubsection{Method 1}
\begin{figure}[htbp]
\begin{center}
\includegraphics[width=9cm,trim=28mm 5mm 13mm 15mm]{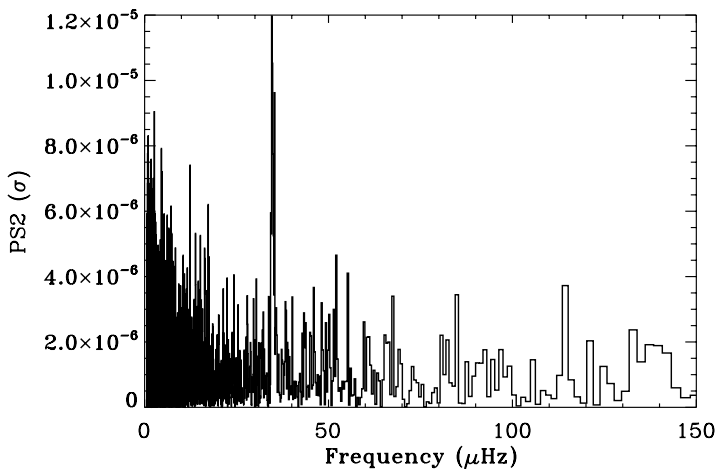}
\caption{Power spectrum of the power spectrum (PS2) between 100 and 10000~$\mu$Hz as a function of the frequency for one simulated star (Pancho) and normalised by $\sigma$ of the PS2.}
\label{fig2}
\end{center}
\end{figure}

As the modes are equally spaced in frequency in the PSD, we compute the fast Fourier transform of a slice between 100 and 10000 $\mu$Hz in the PSD, (called hereafter PS2) and look for the highest peak (see Fig.~\ref{fig2}). 
The large separation is defined as:
\begin{equation}
\Delta \nu = \nu_{n+ 1, \ell} -\nu_{n, \ell},
\end{equation}

\noindent where $\nu$ is the frequency, $\ell$ the degree, and $n$ the radial order of a mode. Between two consecutive orders of a given degree $\ell$ of a mode, we have the $\ell$+1 mode, which is approximately positioned at a distance $\Delta \nu$/2; therefore the periodicity of $\Delta \nu$/2 occurs more often in the PSD. Thus, we assume that the highest peak in the PS2 will correspond to half of the large separation.

Fig.~\ref{fig2} shows an example of a PS2 for a simulated star. We find the highest peak at 34.5 $\mu$Hz, which gives $\langle \Delta \nu \rangle$ = 69~$\mu$Hz. The signal-to-noise ratio of the amplitude of this peak is 6.22~$\sigma$.
Then, in the PSD, we take a box of 600 $\mu$Hz starting at a frequency of 100~$\mu$Hz, calculate the PS2 normalised by $\sigma$, where $\sigma$ is the standard deviation of the PS2 and we look for the peak around $\langle \Delta \nu \rangle$/2 . For this search, we look for the highest peak in the PS2 in the range $\langle \Delta \nu \rangle /2 \pm 10~\mu$Hz. We repeat this by shifting the box in the PSD by 60~$\mu$Hz until 10000~$\mu$Hz. Then, we plot the power normalised by $\sigma$ of the PS2 of the highest peak (or maximum relative power) found in this range as a function of the central frequency of each box. 

\begin{figure}[htbp]
\begin{center}
\includegraphics[width=9cm, trim=28mm 5mm 13mm 15mm]{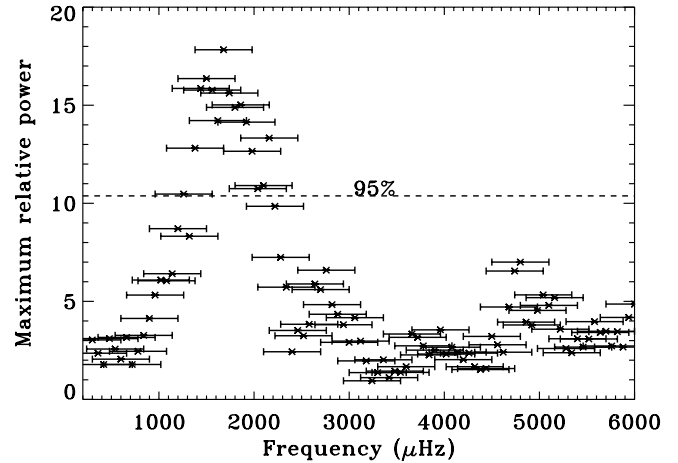}
\caption{Typical maximum relative power of the highest peak in the PS2 around half of the large separation, as a function of the central frequency of the sliding box taken in the PSD for one simulated star (Pancho). The dashed line represents the threshold corresponding to a 95\% confidence level. The error bars represent the size of the boxes taken in the PSD (600~$\mu$Hz in this case). Thus, the range of the p-mode power excess is [940, 2440]~$\mu$Hz.}
\label{pmrs_1}
\end{center}
\end{figure}

An example of this kind of plot is in Fig.~\ref{pmrs_1}. The horizontal error bars correspond to the size of the box of 600 $\mu$Hz. To put some statistical constraints on this search, we calculate the threshold, $s$, corresponding to the probability of finding a peak in the PS2 using the formula \citep{ChaEls2002, GabBau2002}:

\begin{equation}
s = \ln(\Delta T) -\ln P,
\end{equation}

\noindent where $\Delta$ is the size of the box, T is the length of the time-series, and P is the probability. In Fig.~\ref{pmrs_1}, we have plotted the threshold for a probability of 95\%. So we can say that the frequency range where most of the p-mode power is found goes from 937 $\mu$Hz to 2474 $\mu$Hz, which corresponds to the boxes where the relative power of the peak is above the given threshold. Finally, we calculate the mean value of the $\langle \Delta \nu \rangle$ found in each independent box between $f_{\rm min}$ and $f_{\rm max}$ as well as the dispersion giving us the uncertainty on our determination of $\langle \Delta \nu \rangle$.

But as we can have very low SNR stars, we allow the threshold to go down to as low as a probability of 70\% if the higher probabilities are not found.

When we do not find signal above 70\% of confidence level, we change the size of the box of 600~$\mu$Hz and try two different values, 900 and 300~$\mu$Hz, and start the search again. We use a flag {\bf ($box$)} to record the size of the box. 

We have found in some simulated stars that two dissociated regions can be above the threshold, thus two bumps: the correct one and another one due to noise, which gives us a very wide range for the p-mode region. Thus, to determine the correct region, we use again a box of 900 or 300 $\mu$Hz and if we have only one bump in the new analysis, we select the common bump of both analyses. By doing so, we have improved a few results. But if we still have two bumps, we use a flag ($dblpk$) to inform us about the presence of this double bump. 

With the simulated data, we have noticed that the highest peak was not always the one corresponding to half of the large separation. So the first estimation of $\langle \Delta \nu \rangle$ might not be correct and might be due to noise. That is the reason why, if we do not find any signal above 70\% of confidence level in the Maximum relative power, we look for the second highest peak and the third highest peak in the PS2 and do the whole search again. 

We also have a first estimation of $\nu_{\rm max}$, the frequency of maximum amplitude, by taking the frequency of the highest Maximum relative power.

Once we have determined $\langle \Delta \nu \rangle$ and we have an estimation of $\nu_{\rm max}$, we verify our results by using scaling laws \citep{stellochap2009}, which establish the following relation:

\begin{equation}
\langle \Delta \nu \rangle=135 \times \Big( \frac{\nu_{max}}{3050} \Big)^{0.8}.
\end{equation}

\noindent If the $\langle \Delta \nu \rangle$ found with our method does not fall within an uncertainty of 20\% around the theoretical value, we use a flag. 

As an alternative, we can run this code a second time with an estimation of $\langle \Delta \nu \rangle$ as an input and by constraining our search for a peak in the PS2 around this theoretical value $\pm$15~$\mu$Hz. 

The output of this part of the pipeline is the value of the large separation and the frequency range of the p-mode bump with a given confidence level.

\subsubsection{Method 2 }
The second method to look for the large separation is based on the estimation of some parameters with spectroscopic observations. From the fundamental stellar parameters: parallax, effective temperature, apparent magnitude, and log $g$ or as in the Kepler stars: effective temperature, log $g$, and radius, given in the KIC \citep[Kepler Input Catalog,][]{2005AAS...20711013L}, we use the scaling laws from \citet{1995A&A...293...87K} to estimate the range in frequency for the p-mode excess power and the large separation.
With these initial values, we look for $\langle \Delta \nu \rangle$ in the power spectrum (PS) of the data as well as in some average spectra by taking  several subseries in the data. We can choose the number of average spectra.

\begin{figure}[htbp]
\begin{center}
\includegraphics[width=9cm, trim=28mm 5mm 13mm 15mm]{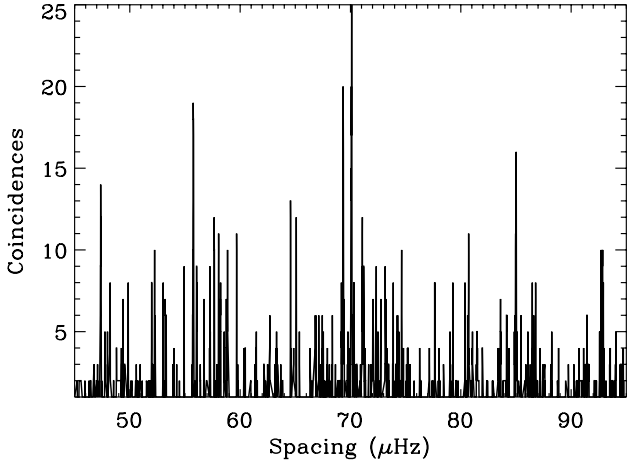}
\caption{Result for the search for the spacing in one month of data
for one of the simulated stars (Pancho). The highest peak
corresponds to the large separation present in the data.}
\label{pmrs_2}
\end{center}
\end{figure}

The search is done by computing the PS of a short slice (typically 900~$\mu$Hz) of the PS where the p-mode excess power appears \citep{2002A&A...396..745R}. The search for the spacing is done iteratively by trying different values in a 50~$\mu$Hz-range around the estimated spacing and with a step of 0.01 $\mu$Hz. We try to find if there is any signal above 1.5 times the rms of the power spectrum at any of the bins spaced by $\langle \Delta \nu \rangle$. To evaluate the significance of the peaks found and to avoid binning effects, this procedure is repeated 50 times and by continuously shortening the length of the 900~$\mu$Hz-slice, down to 1$\times \langle \Delta \nu \rangle$. The coincidences of spacing among the peaks found in the 50 trials are then registered (See Fig.~\ref{pmrs_2}).

From this search we obtain for the whole spectrum as well as for the averaged spectra a possible value of $\langle \Delta \nu \rangle$. We also fit the histogram of coincidences with a Gaussian profile. The width of the latter gives the error bar ($\sigma$) associated to the value obtained for the spacing. Then, we take the mean value of all the $\langle \Delta \nu \rangle$ found in each data series, typically 5 for a month of data. The values that are 3~$\sigma$ above the mean value are rejected and a new mean value with its error is estimated. This process is repeated 3 times and the result with its error is taken as the large separation for the analysed data set. However, if the data have a poor SNR or if the fundamental stellar parameters are badly determined, it can lead to the wrong spacing, and it is very difficult to identify this result as incorrect.



In the possible case for which the fundamental stellar parameters are ill-determined, we have the possibility to run all the pipeline a second time using as an estimation for the p-mode frequency range and for the large separation, the results obtained blindly by Method 1 described above.

\subsubsection{Cross-checking the two  methods}

To cross-check the large separations obtained with the two methods, we follow three steps:

First step: We run both methods,  Method 1 based on the blind estimation of the initial parameters,  and Method 2, with the initial parameters obtained from the scaling laws. We obtain some results that are compatible between both methods and some that are different.

To cover the possibility to have selected a wrong region for the p-mode range in the blind way of finding it in Method 1 or the possibility to have a bad determination of the fundamental parameters in Method 2, we implement two more steps.

Second step: Method 2 is run using the blind estimations from Method 1 and the new results of Method 2 are compared with the results from Method 1.

Third step: Method 1 is run using the scaling law estimations from Method 2  and the new results of Method 1 are compared with the results from Method 2.

As explained in Sect.~2.2.1, we calculate the threshold for detecting a peak in the PS2 for a given confidence level. We say that we are very confident when we are above a threshold of 85\% (confidence level of 2), less confident between 70 and 85\% (confidence level of 1), and not confident at all below 70\% (confidence level of 0). 

To select the most reliable results, we dismiss the step in which we obtain the highest number of common stars with a confidence level of 0. Among the 2 remaining steps in which we remove the stars with the confidence level of 0, we choose the step where we have the highest number of coincidences.


As a fourth step, to try to recover the spacing of more stars, we look at the common stars in the two comparisons in which we have obtained the lower number of stars with zero confidence level. For example if among 300 stars we have 170 stars with common results with step 1 (with a confidence level of 1 and 2) and in step 2 there are 195 of them, we check if the first 170 common stars from step 1 are included in the 195 from step 2. If only 150 stars are the same between step 1 and 2, we accept the other 20 as correct results and we add them to the 195 obtained from step 2. Finally, we will have 215 common stars.

\subsection{Background Fitting}


The background of the spectrum is modeled with three components:
\begin{enumerate}
\item white noise to model the photon noise (W);
\item Harvey's model which reproduces the convective contribution to the background \citep{1985ESASP.235..199H};
\item a power law modelling activity effects and low-frequency trends.
\end{enumerate}
Our model has then 6 free parameters:
\begin{equation}\label{eq:bgmodel}
 B(\nu)=W+\frac{A}{1+\left(2\nu/\nu_c\right)^\alpha}+a\nu^{-b}.
\end{equation}
All of the parameters have to be positive, which is a constrain in the fitting code. The model is fitted to the spectrum over the domain $[\nu_1,f_{\mathrm{min}}]\cup[f_{\mathrm{max}},\nu_{\mathrm{cut}}]$ using a classical maximum-likelihood estimator. 

The parameters $f_{\mathrm{min}}$ and $f_{\mathrm{max}}$ are the bounds of the interval where p-mode power excess has been detected and returned as an output of the package PMRS. $\nu_{\mathrm{cut}}$ is the cut-off frequency and $\nu_1$ the lowest considered frequency (in practice, we consider  $\nu_1=1\mathrm{\:\mu Hz}$). In other words, we simultaneously fit the full spectrum after having excluded the p-mode region.

We use as an initial guess for $W$ the average of the power spectrum around the cut-off frequency, where the photon noise is expected to dominate the other components. To find guesses for $a$ and $b$, we perform a log-log regression in the low-frequency domain. The only input guess requested by this package is $\nu_c$. The value of the spectrum around the frequency $\nu_c$ is then used to extrapolate a guess for $A$. We have typically used values in the range 100--1000$\mathrm{\:\mu Hz}$ for the guess of $\nu_c$. By using this set of guesses, a first fit is performed where $\alpha$ is kept fixed with a value of 2. The fitted values are then used as guesses for a second fit where all of the parameters have been let free.

We have noticed with our benchmark of tests that this routine is robust and stable. The final results depend indeed only weakly on the exact value used for the input guess of $\nu_c$.


\subsection{Characterisation of the P-Mode Envelope (Bolometric Amplitude per $\ell$=0)}


Once we have fit the background with the package BF, we subtract it from the PSD as we need to estimate the amplitude of the modes without the contribution of the stellar noise. Then, we smooth the new PSD using a sliding window of $q$~$\times \langle \Delta \nu \rangle$, where $q$=1, 2 or 3. We obtain the envelope of the power spectrum. We fit the result with a Gaussian function, which gives the maximum power, $P_{\rm max}$ in $\mathrm{ppm}^2 \mu \mathrm{Hz}^{-1}$ and the frequency position, $\nu_{\rm max}$. We also recover the error estimate from this fitting, which gives us the standard deviation related to both coefficient, $P_{\rm max}$ and $\nu_{\rm max}$.


\noindent To convert this power into amplitude, $Ampl_{max}$, we use the following formula as we have a single-sided PSD:

\begin{equation}
Ampl_{\rm max}=\sqrt{2\times P_{\rm max} \times q \times \langle \Delta \nu \rangle}.
\end{equation}

\begin{figure}[htbp]
\begin{center}
\includegraphics[width=9cm, trim=28mm 5mm 13mm 15mm]{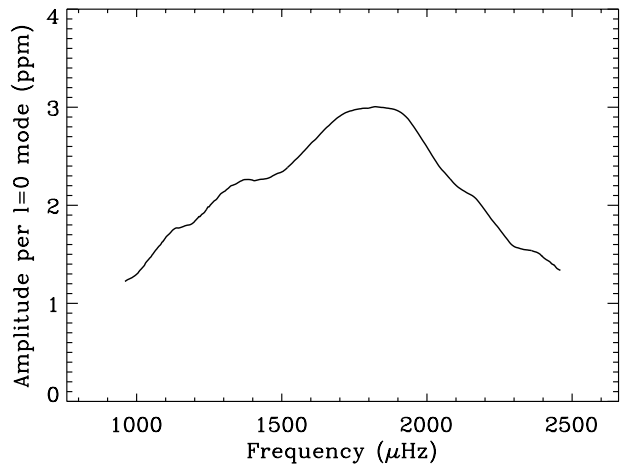}
\caption{Typical bolometric amplitude of the mode $\ell$ = 0 as a function of the frequency for one simulated star.}
\label{fig5}
\end{center}
\end{figure}

\noindent Then, we want to calculate the rms maximum amplitude per radial mode or the bolometric amplitude, $A_{\rm max}$. As we use the method described in \citet{2008ApJ...682.1370K}, the formula for the amplitude is the following: 

\begin{equation}
A_{\rm max}=Ampl_{\rm max}/\sqrt{ 2 \times q \times c}= \sqrt{(P_{\rm max} \times \langle \Delta \nu \rangle)/c},
\end{equation}

\noindent where $c$ is related to the spatial response to the observations of the modes of degree $\ell$ = 0 to 4 and to the type of observations (intensity or velocity). For the simulated stars, we took c=3.16 as given in \citet{2008ApJ...682.1370K}. However, the correction to obtain the bolometric amplitude per radial mode depends both on the instrument and the star ($T_{\rm eff}$), thus it should be calculated for each star and for a given instrument. This has been done by \citet{2009A&A...495..979M} for the CoRoT mission and should be done for the Kepler mission as well. Finally, we can plot the bolometric amplitude per radial mode in the frequency range found in the package PRMS, which is shown in Fig.~\ref{fig5}.

From now on, when we talk about the maximum amplitude of the mode, we will refer to the rms maximum amplitude.

\subsection{First Determination of the Mass and the Radius}

We assume here that we do not have any information on the radius and the mass of the star beforehand, which is usually the case. Instead of using the scaling laws \citep{1995A&A...293...87K, 2003PASA...20..203B} to estimate $\langle \Delta \nu \rangle$ and $\nu_{\rm max}$, we use them as input. We solve the two equations with two unknown variables: the radius of the star $R'$ and its mass $M'$. We can calculate them according to formulae (9) and (10) from \citep{1995A&A...293...87K}.

\begin{equation}
\frac{R'}{R_\odot}=\Big( \frac{135}{\langle \Delta \nu \rangle}\Big)^2 \Big(\frac{\nu_{\rm max}}{3050} \Big) \Big(\frac{T_{\rm eff}}{5777} \Big)^{1/2}\\
\end{equation}

\begin{equation}
\frac{M'}{M_\odot}=\Big(\frac{135}{\langle \Delta \nu \rangle} \Big)^4 \Big(\frac{\nu_{\rm max}}{3050} \Big)^3 \Big(\frac{T_{\rm eff}}{5777} \Big)^{3/2}. \\
\end{equation}

Concerning the uncertainty on these values, we have to take into account the error bars on $\langle \Delta \nu \rangle$, $\nu_{\rm max}$. This leads us to uncertainties of about 10\% and 20\% respectively for the radius and the mass. These results depend on the validity of the scaling laws. It has been recently shown, thanks to the CoRoT mission, that they are closely followed for red giants \citep{2009arXiv0906.5002H, 2009Natur.459..398D} and one of the first goals of the asteroseismic investigation of the Kepler mission will be to check their validity.

This package gives a first estimation of the mass and radius while the package described in the next section to estimate the radius is more accurate but more time consuming.


\subsection{Guessing Parameters and variation of the large separation with frequency}

 In this section, we describe how the guess-parameter table is calculated, which is needed to perform the peak fitting of the individual modes. The guess frequencies thus obtained provide a first estimation of the frequency of each ($\ell$, $n$) mode and the  large-separation variation with frequency.

The power spectrum is corrected from the background obtained by the package BF and smoothed by a factor of 2\% of the estimated mean large separation $\langle \Delta \nu \rangle$ (Sect.~2.2).
The guess table is constructed over the frequency range $[f_{\rm min},f_{\rm max}]$ where p-mode power excess was found (see Sect.~2.2).  The frequency bin of the highest peak in the power spectrum around the estimated $\nu_{\rm max}$ (Sect.~2.4) is found. Then, the consecutive mode frequencies $\nu_{g,i\pm 1}$ are determined by searching for the highest peaks within frequency windows of $(\nu_{g,i} \pm  \Delta \nu_{i}) \pm 0.15 \Delta \nu_i$, where the large separation $\Delta \nu_i$ is continuously updated based on the previously measured frequencies.


The guesses of the Lorentzian-mode amplitudes are calculated from the maximum power corresponding to each guess frequency.



\begin{figure}[htbp]
\begin{center}
\includegraphics[width=9cm, trim=0.5cm 0.2cm 0.5cm 0cm]{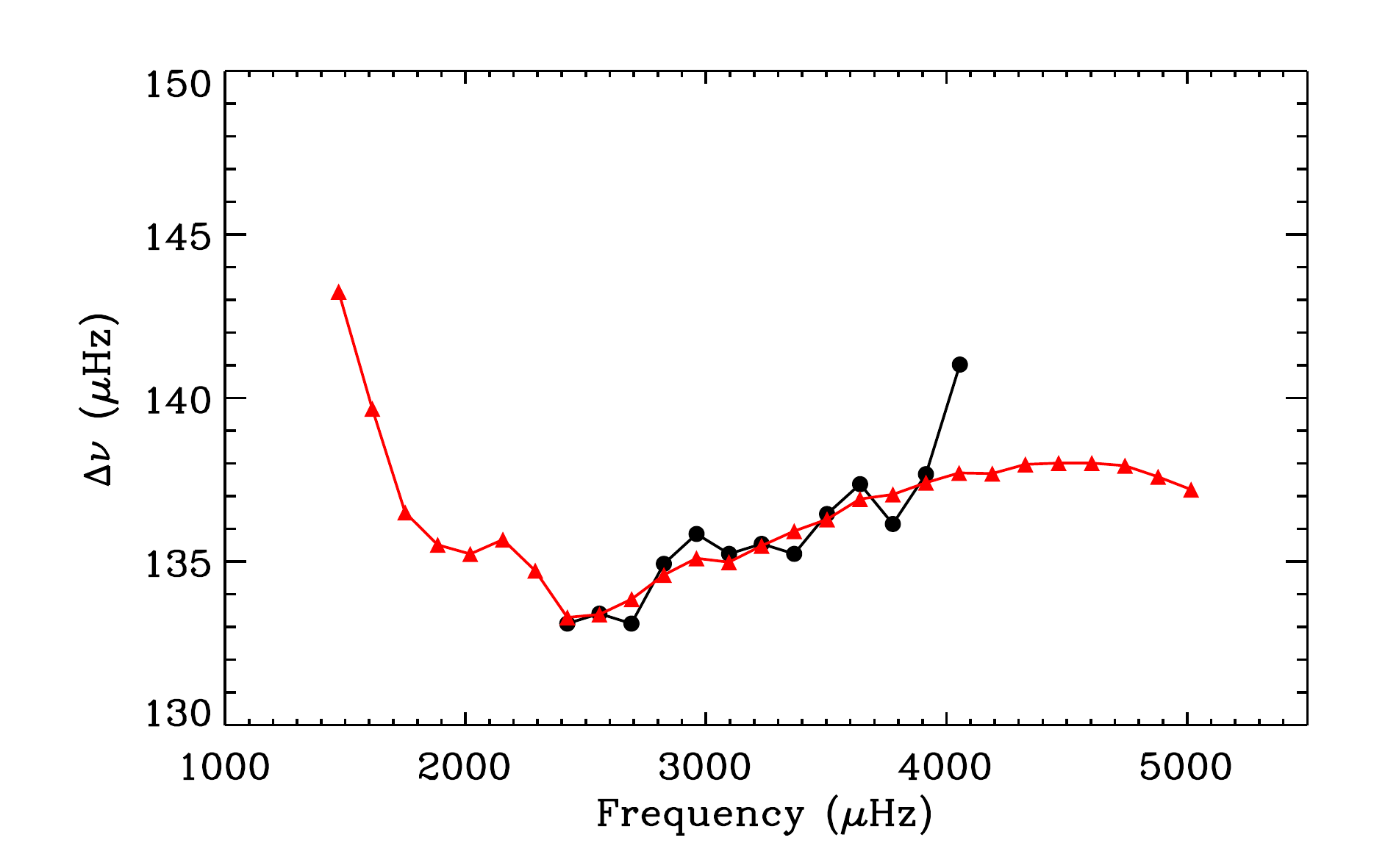}
\caption{Frequency dependence of the large separation of the group of even modes $l=0-2$ measured from 30 days of one simulated asteroFLAG star (Boris). The triangles represent the input values, and the circles the results from the pipeline.}
\label{var_dnu_boris}
\end{center}
\end{figure}

A first estimate of the frequency dependence of the large separation $\Delta\nu(\nu)$ is thus obtained. An example for 30 days of a simulated asteroFLAG star is shown in Fig.~\ref{var_dnu_boris}.

\section{Radius from a grid of stellar models}

Once we have robust estimates of $\langle \Delta \nu \rangle$, 
$f_{\rm min}$ and $f_{\rm max}$ (see Sect.~2.2),  we couple this 
information with the available atmospheric parameters
to estimate the 
radius of the star using stellar models.
In the specific case of Kepler data, the atmospheric parameters that we use
and that are available to
us from the KIC
are
$\log g, [M/H]$, and $T_{\rm eff}$.
To determine the global parameters, we compare the observations 
($\langle \Delta \nu \rangle$, $\log g, [M/H]$, and $T_{\rm eff}$)
with model observables from 
stellar evolution and pulsation codes.  
We use the Aarhus Stellar Evolution Code (ASTEC) 
coupled to an 
adiabatic pulsation code (ADIPLS) \citep[][]{2008Ap&SS.316...13C, 2008Ap&SS.316..113C}.  
These codes need as input the stellar parameters of mass, age, chemical
composition 
and mixing-length parameter, and return the stellar model interior profiles, 
global parameters, such as radius and effective temperature, and 
the frequencies of the oscillation modes.

The parameters that best describe the observables are obtained by 
minimising a $\chi^2$ function; 
\begin{equation}
\chi^2 = \sum^M_{i=1} \left ( \frac{y_i - B_i}{\epsilon_i}\right )^2,
\end{equation}
where $y$ and $B$ are the $i=1,2,...,M$ observations and model observables.
The Levenberg-Marquardt algorithm is used for the optimization, and this
incorporates derivative information to guess the next set of parameters
that will reduce the value of $\chi^2$.  Naturally, an initial guess
of the parameters is needed and these are obtained from a small grid 
of stellar evolution tracks ($T_{\rm eff}, \log g$)
where we assume 
$\langle \Delta \nu \rangle$ is proportional to the mean density of the star.

\begin{figure}
\includegraphics[width=0.48\textwidth]{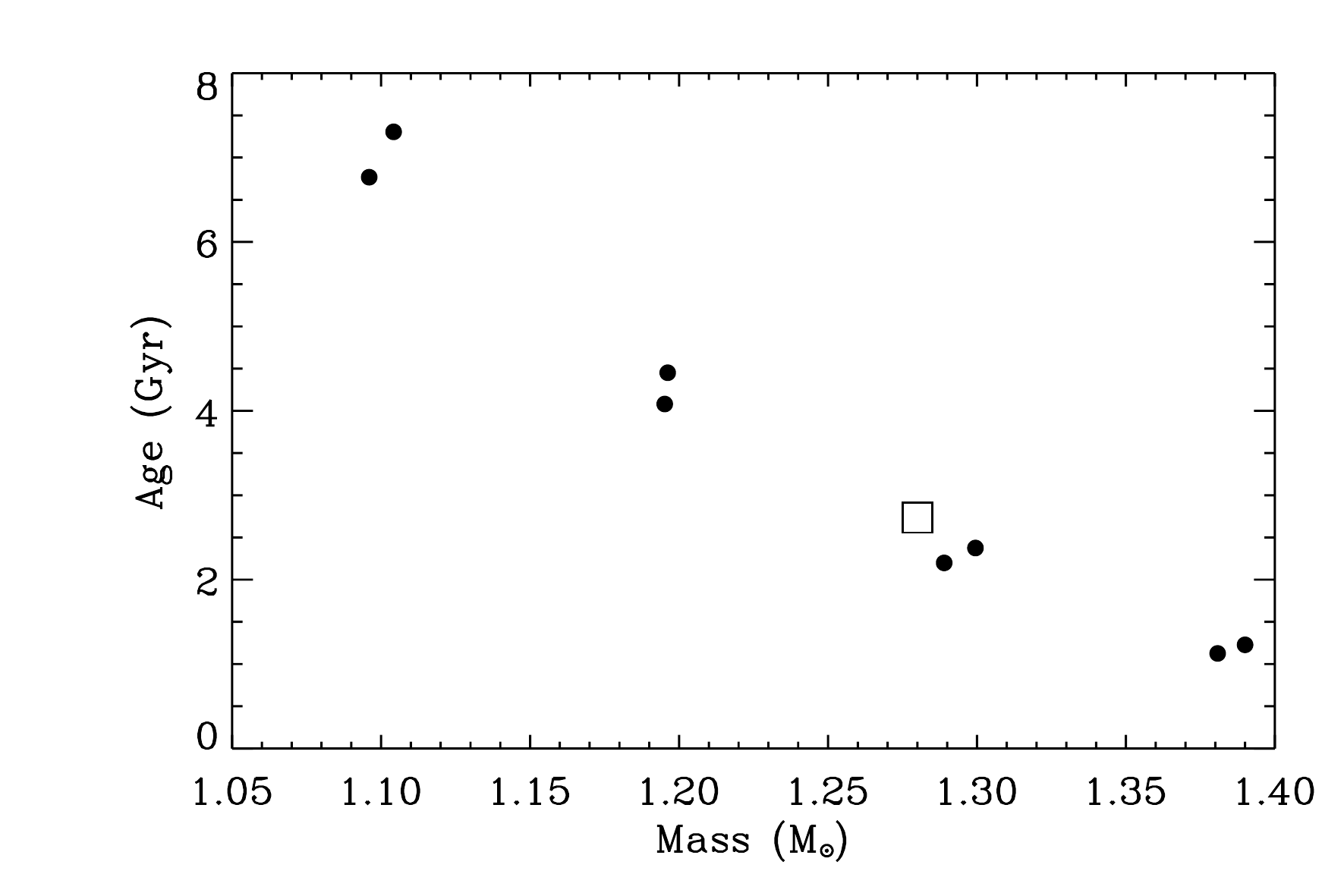}
\caption{The fitted masses and ages for a simulated star. The filled circles show the masses and ages derived from our fitting procedure starting from eight different initial guess values.
The input value is denoted by the square.\label{fig:orl1}}
\end{figure}

\begin{figure}
\includegraphics[width=0.48\textwidth]{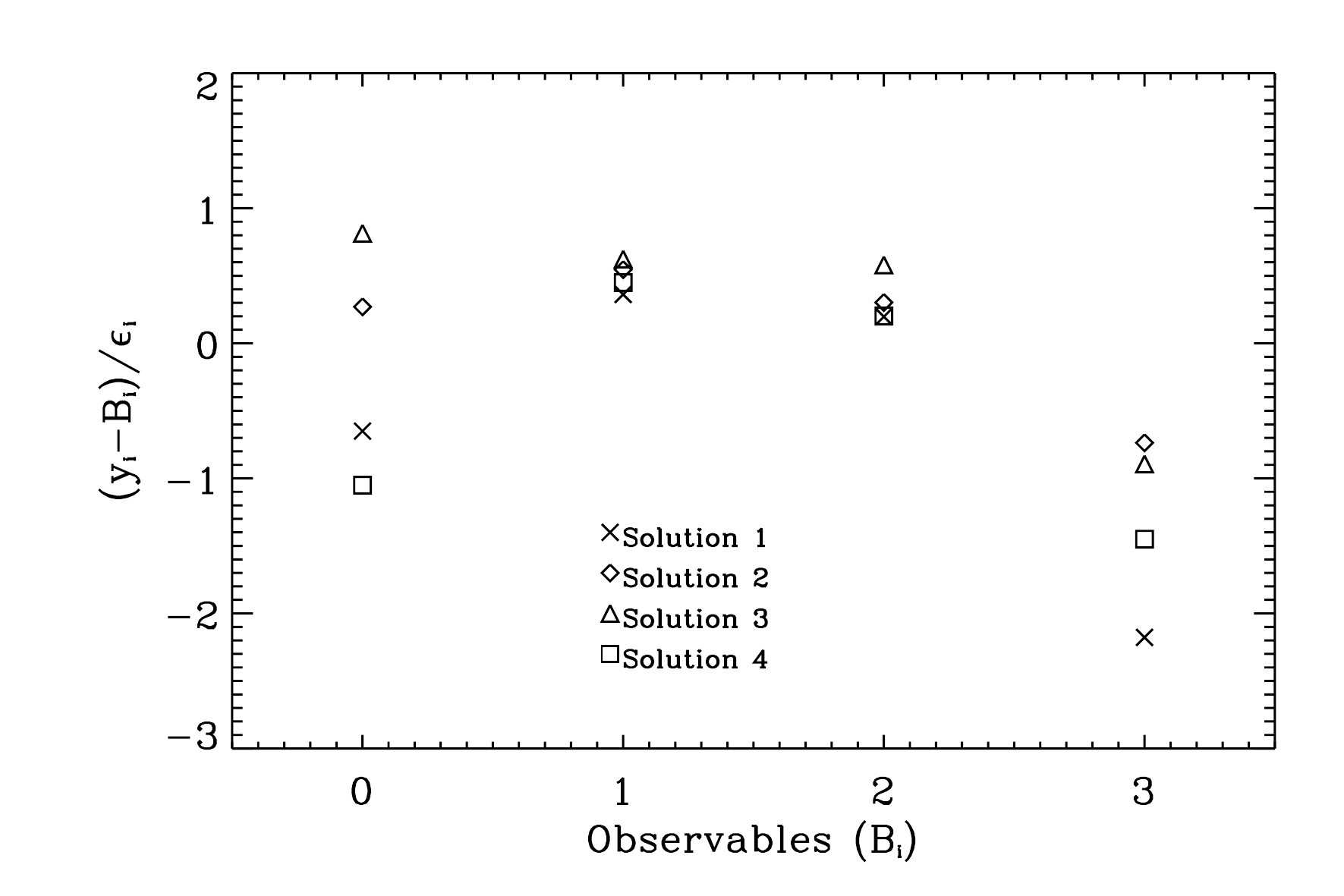}
\caption{The residuals between the observations and model observables for 
four of the models from Fig.\ref{fig:orl1}. From left to right, the observables are: $T_{\rm eff}$, $\log g$, $[M/H]$, and $\Delta \nu$.\label{fig:orl2}}
\end{figure}

Because there are few observations and just as many parameters, there are
inherent correlations between mass, age, and metallicity. 
To help avoid local minima problems, we minimise the $\chi^2$ function 
beginning at several initial guesses of the parameters (mainly varying
in mass and age), and these initial guesses are estimated from the grids. 
For each star, we therefore obtain several sets of parameters with a 
corresponding $\chi^2$ value that match the observations as best as possible.
To highlight the correlations we obtain, 
Fig.~\ref{fig:orl1} shows several
values of mass and age that adequately match the observations 
of a simulated star. 
The square symbol highlights the true input values of these models.
Fig.~\ref{fig:orl2} shows the residuals (observation - observable) 
from 4 of the best-fitting models.
  It can be understood quickly from these figures, that
there are difficulties with extracting both the mass and the age of the stars.

 
Luckily, the radius parameter
is highly dependent on the set of observations. This means that although
we obtain quite a scatter in mass and age for each of the simulated stars, 
we do
obtain robust estimates of the radius.

The radius is calculated as the average value from the best-fitted models 
where the $\chi^2$ value
is below 3.9$^2$.  
The error is defined as one sixth of the difference between
the maximum and minimum radius.
The value of six stems from the ``error'' being defined as half of the scatter 
in values, and then we divide by 3 to produce an estimate of the 1 $\sigma$ 
uncertainty, assuming that the range of values spans 3.9~$\sigma$. 
The choice of dividing by 3 instead of 3.9 is to account for the low number
of models that we are using to estimate the paramaters.


\section{Results}

\subsection{Simulated data}
In this work, we have used simulated data generated within the AsteroFLAG team for hare-and-hounds exercises \citep{2008AN....329..549C} dedicated to prepare the Kepler mission. We have also applied our pipeline to data from the Aarhus simulator \citep{2004SoPh..220..207S}. These simulated stars are based on the information available in the KIC of the stars that will be first observed by Kepler during the survey phase.

\subsubsection{AsteroFLAG hare and hounds}

We have decided to work on three stars (or ``cats'') simulated by the AsteroFLAG team: Arthur, Boris, and Pancho. For each of them, we have 2 different rotation rates (low and high), 3 angles of inclination (0, 30, and 60$^\circ$), and 4 magnitudes (from 9 to 12).  From these 72 simulated stars with different fundamental parameters, 2592 independent spectra of one month have been obtained (36 for each star).
These 2592 spectra have been analysed independently with the methods described above for the package PMRS and the results have been compared following our cross-checking methodology (Sect.~2.2.3) for estimating $\langle \Delta \nu \rangle$.\\

Table~\ref{tab1_stat_cats} shows the statistics of the common results found for all the 2592 spectra according to the type of star. We can see that by exchanging our results in steps 2 and 3, the number of common results increases. For a given step, if we have a high number of common stars with a confidence level of 0, it means that the common results could be incorrect. This is due to the fact that if there is some signal (real or not) and if both methods search around the same region, they should find almost similar results. 

\begin{table*}[h]
\caption{Number of common results for the asteroFLAG and Aarhus/asteroFLAG stars, for the 3 steps of our strategy and the corresponding number of results with a 0 confidence level.}
\begin{center}
\begin{tabular}{ccccccc}
\hline
\hline
Star & Step 1 &Level 0& Step 2 &Level 0& Step 3&Level 0\\
\hline
Arthur & 182 & 11 & 223 & 10 & 294 & 62\\
Boris & 300 & 5 & 305 & 11 & 531 & 135\\
Pancho & 231 & 6 & 259 & 9 & 310 & 273\\
Aarhus/asteroFLAG& 33 & 3 & 75 & 4& 78& 36\\
\hline
\hline
\end{tabular}
\end{center}
\label{tab1_stat_cats}
\end{table*}%

\subsubsection{Aarhus/asteroFLAG simulator stars}

We have also tested our pipeline on 176 stars simulated with the Aarhus/asteroFLAG simulator. These stars have magnitudes from 7 to 11 and they are one-month time-series.

Once again we have analysed the common results for the different steps of our methodology of the PMRS package (Table~1). For step 1, we have 33 results in common with 3 results with level 0. For step 2, we have 75 results in common and 4 level 0. Finally, for step 3, we have 78 common results with 36 level 0.

\subsubsection{Strategy to select reliable results}

The strategy adopted in this work is to look at the number of common results in each step described previously and among them, to take into account the number of results with a confidence level of 0. We have to make a trade-off to have the highest number of common stars with the lowest number of zero-confidence level stars.

We recap here how we select the best step. First, we remove the step where the number of stars with a confidence level of 0 is the highest. Then, we choose the step with the highest number of common stars and lowest number of stars with 0 confidence level. We select only the results where the confidence level is equal to 1 and 2, so we subtract the number of results with a zero-confidence level. Finally, we add up the stars found in the 2 best steps as all the common results  might not be  included in the common results of the other step.

By comparing the results to the input, a tolerance of 5\% of the real value for $\langle \Delta \nu \rangle$ is accepted. If we are within this error, we say that the result is ``correct''.


For instance, for the case of Pancho (Table~\ref{tab1_stat_cats}), we notice that for steps 1 and 2, the number of confidence level of 0 is quite close: 6 and 9.  But the number of common results is higher for step 2, 259 compared to 231 for step 1. So we choose the common results of step 2. When we remove the results with a confidence level of 0, this gives 250 confident results. By applying it to all the stars, we select: 25\% of the common results with Arthur, 34\% for Boris, 29\% for Pancho, and 40\% for the Aarhus-asteroFLAG simulated stars. Among them, we know that there are respectively 9\%, 2\%, 11\%, and 6\% of incorrect results.

Finally, according to the last step of our methodology, if we add the stars that are in common in step 1 in the case of Pancho and that do not appear in the common stars in step 2, we obtain 293 stars for which we are confident. Thus we would have 29\% of common results for Arthur, 38\% for  Boris, 34 \% for Pancho, and 40\% for the Aarhus-asteroFLAG simulated stars, with respectively  16.8\%, 2.8\%, 13.3\%, and 6\% of incorrect results.


Now, if we look at the results in terms of magnitude (Table~\ref{tab2_stat_cats}, first half) we notice that before applying the last step, for a magnitude V=9, we manage to have 87\% of common results, with only 0.5\% of incorrect results. Then, for V=10, the number of common results decreases to 25.6\% with 10.8\% of bad determinations. Finally, with higher magnitudes, the results are not convincing as we have more than 50\% of wrong determinations.  So we can say that our pipeline works well for stars with V$<$11.

After applying the last step (Table~\ref{tab2_stat_cats}, second half), the results are the following: for V=9, we have 91.5\% of common results, with only 0.7\% of incorrect results, for V=10, we have 33.3\% of common results with 14.3\% of bad determinations.

\begin{table*}[htdp]
\caption{Number of common results for all the steps of our strategy with the errors (incorrect results) as a function of the magnitude. The second half of the table shows the results when we add the common results of the two best steps. In the parentheses, we put the percentage of incorrect results.}
\begin{center}
\begin{tabular}{ccccccccc}
\hline
\hline
Star / V & 9 & Error& 10&Error& 11 &Error& 12 & Error\\
\hline
Arthur & 176 & 2 (1\%)   & 25 & 5 (20\%)  & 5  & 5 (100\%) & 7 & 7 (100\%)\\
Boris  & 209 & 0 (0\%)   & 89 & 2 (2\%)   & 3  & 1 (33\%)  & 2 & 2 (100\%) \\
Pancho & 179 & 1 (0.5\%) & 52 & 11 (21\%) & 10 & 8 (80\%)  & 9 & 8 (89\%) \\
\hline
\hline
Arthur & 182 & 2 (1\%)  & 39  & 14 (36\%)  & 14  & 12 (85\%)  & 14 & 14 (100\%)\\
Boris  & 214 & 0 (0\%)  & 107 &  3 (3\%)   & 7   & 2  (28\%)  & 4 & 4 (100\%) \\
Pancho & 197 & 2 (1\%)  & 70  & 14 (20\%)  & 13  & 11 (84\%)   & 13 & 11 (84\%) \\

\hline
\hline
\end{tabular}
\end{center}
\label{tab2_stat_cats}
\end{table*}%

\begin{figure}[htbp]
\begin{center}
\includegraphics[width=9cm]{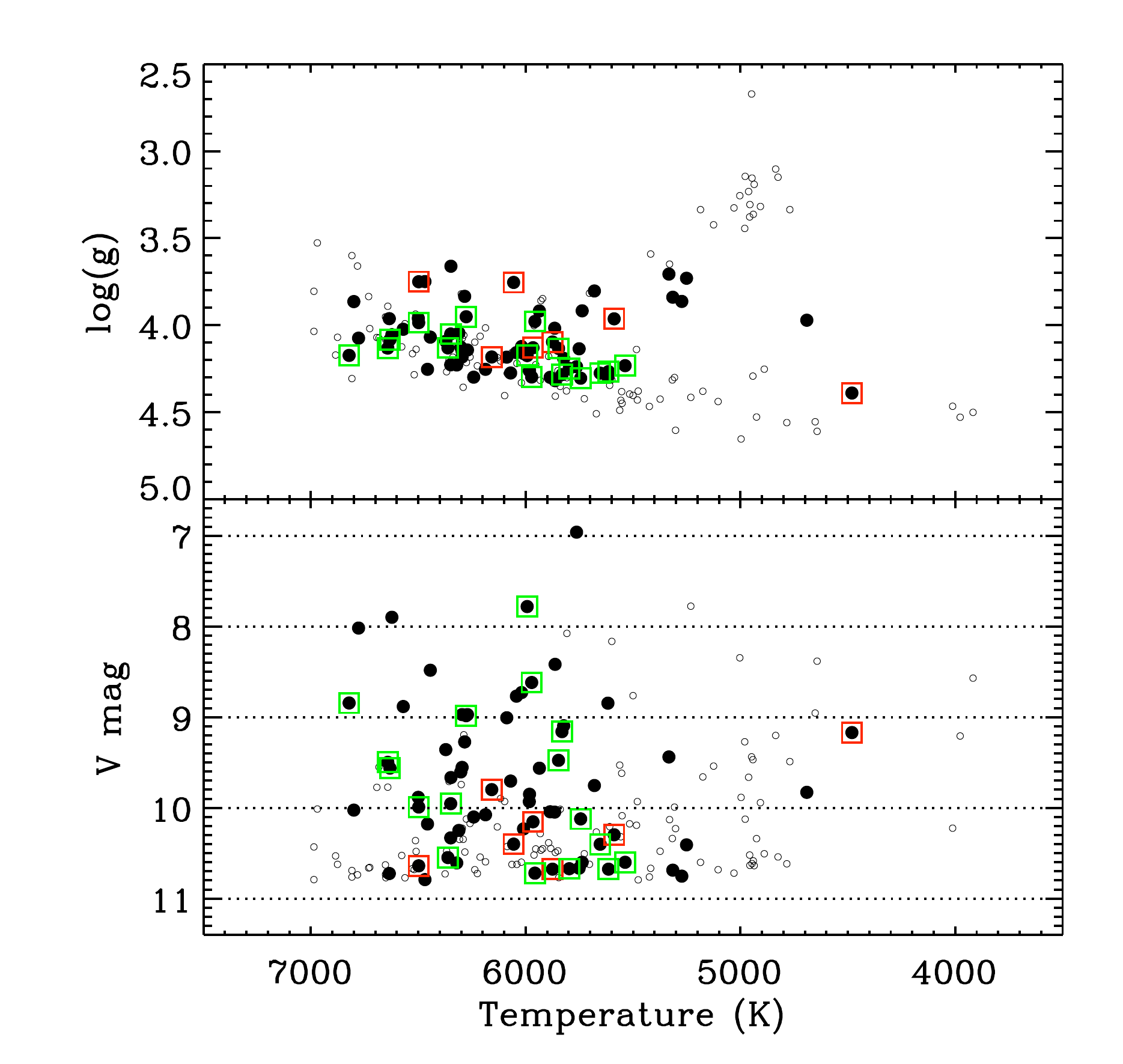}
\caption{Distribution of the correct results obtained with the simulated stars from the Aarhus-asteroFLAG simulator in term of V  magnitude, $T_{\rm eff}$, and log $g$. The open circles represent the 176 simulated stars. The full black circles represent the stars for which the correct value of the large separation was obtained. The squares are the stars selected to determine the radius: the green squares are the converging models and the red squares are the diverging models.}
\label{survey_res}
\end{center}
\end{figure}

For the simulated stars of Aarhus/asteroFLAG, this strategy leads to 71 common results out of 176 (40\%), among which 67 results are correct. Applying the last step does not change the result. Fig.~\ref{survey_res} represents the distribution of correct results in terms of magnitude, $T_{\rm eff}$, and log $g$. We notice that for magnitudes lower than 8, 75\% of the stars are retrieved. For 8$\le$V$<$9, we have a good estimation of $\langle \Delta \nu \rangle$ for 63.1\% of the stars, for  9$\le$V$<$10, 48.9\% of the stars, and for V$\ge$10, 27.1\% of the stars. These results are quite close to what we have obtained with the cats. 

\begin{figure*}[htbp]
\begin{center}
\includegraphics[width=15cm]{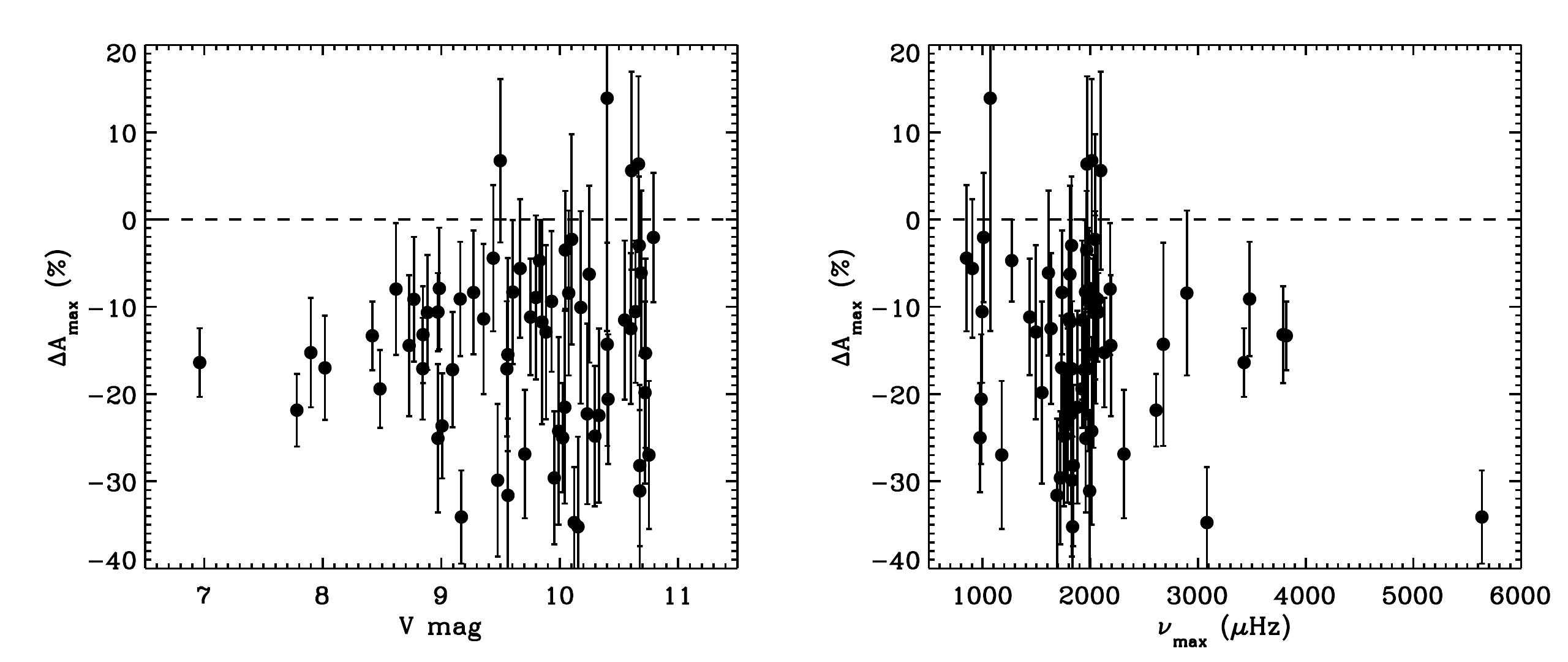}

\caption{Relative difference between the maximum amplitude calculated with the \emph{A2Z} pipeline (using q=3) and the input values of the Aarhus-asteroFLAG simulated stars as a function of the magnitude (left panel) and the frequency of the maximum amplitude (right panel). We overplotted the error bars for each estimation.}
\label{comp_maxampl}
\end{center}
\end{figure*}

We have applied the other packages, BF and BAL0, to these 67 stars and estimated the maximum amplitude per radial mode. Fig.~\ref{comp_maxampl} shows the difference between the maximum amplitude estimated (using a smoothing of 3$\times \Delta \nu$) and the input. We clearly notice that for most of the stars, we underestimate the amplitude per radial mode. This can be due to the smoothing of the spectrum and maybe to the background fitting. We notice that the Gaussian fit is most of the time below the p-mode envelope. There is not any correlation between the correct estimation of the amplitude and the magnitude of the star (Fig.~\ref{comp_maxampl}, left panel), or with the position of the maximum amplitude in frequency (Fig.~\ref{comp_maxampl}, right panel). For 73\% of the stars, we manage to retrieve the maximum amplitude per radial mode within 3~$\sigma$.
 We have also calculated these amplitudes with different values for $q$. The bias varies between -10.8~$\pm$9\% and -14.6~$\pm$8\%. For instance, with a smoothing of 1$\times \Delta \nu$, we obtain that 81~\% of the results are within 3~$\sigma$. 


\begin{figure}
\includegraphics[width=0.48\textwidth]{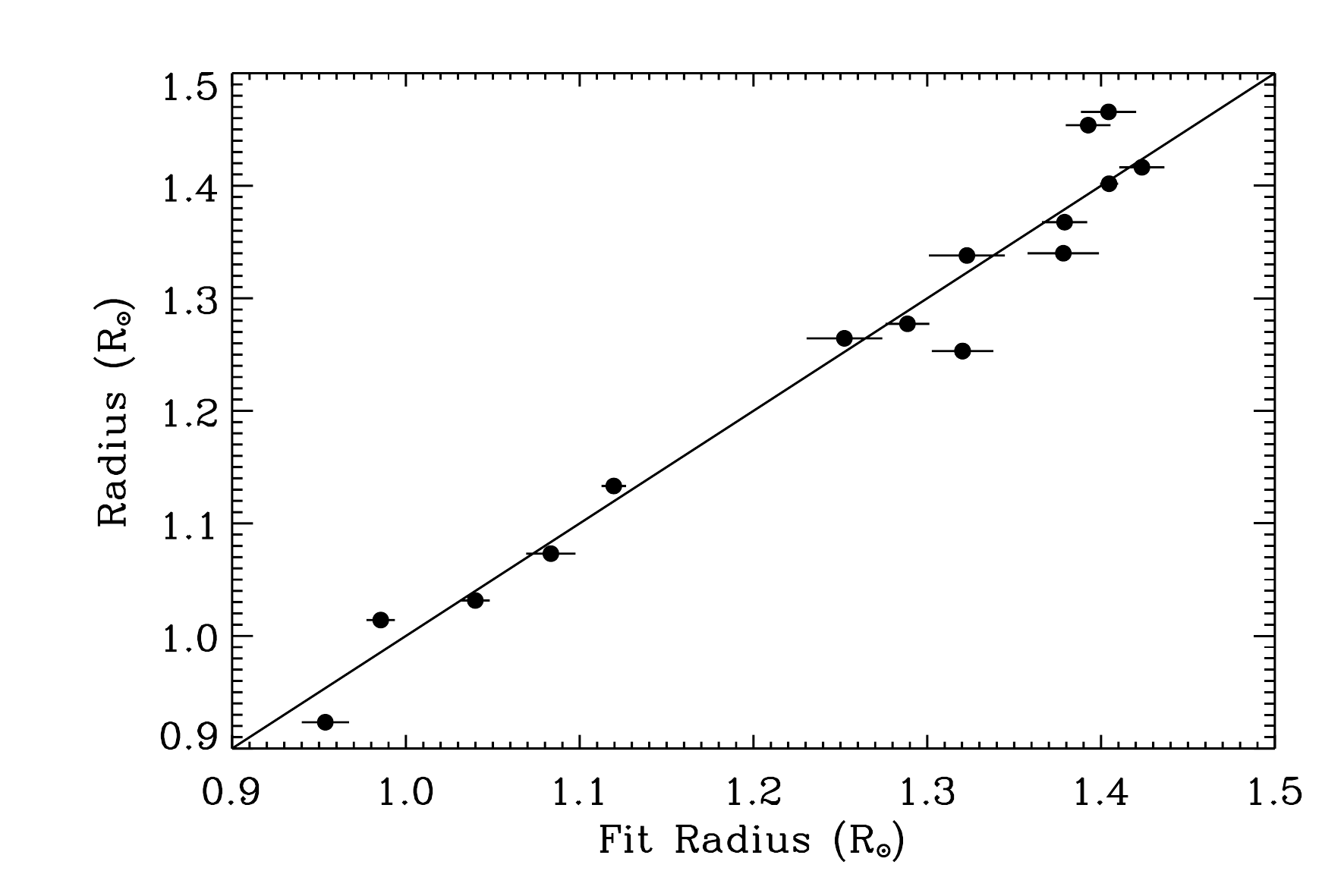}
\caption{The true radius versus the fitted input radius for a selection of the 
simulations.  The continuous line is the x=y line.
\label{fig:orl3}}
\end{figure}

For a few tens of the survey stars (squares in Fig.~\ref{survey_res}), we have calculated the radius following the method described in Sect.~3. Fig.~\ref{survey_res} shows the models that converged with our method (green squares) as well as the models that do not converge (red squares). We selected only a few tens of stars for time and computer limitations. The selection was also done such as to have a panel of different stars, specially with a wide range for the mean large separation. The results are given in Table 3.  We also give the difference between the true input value and the value obtained with the pipeline and among 18 stars, we obtain the radius within 3~$\sigma$ for 13 stars.
Fig.~\ref{fig:orl3} shows the input values of the radius for the simulated data versus the fitted values and their errors bars obtained with the pipeline.
The continuous line is the x=y line.
The figure shows that the determination of the
radius from the pipeline is a rather good determination of the true input values within a few\%.
We note that there are some stars that were not estimated, and the
stellar models most likely did not converge due to unphysical sets of parameters
as input.
Some more work is needed in the code, so that the
automatic pipeline will give results for $>95$\% of the stars.  At the moment the
success rate is $\sim$70\%.

\begin{table*}
\begin{center}
\caption{Radius ($R_F$) from stellar modelling for some Aarhus/asteroFLAG simulated stars\label{table:radius_simul} compared to the simulation input ($R_M$).}
\begin{tabular}{lcccccccccccc}
\hline\hline
KIC ID &$R_{\rm F}$ ($R_{\odot}$) &$\sigma(R)$ & $R_{\rm M}$ ($R_{\odot}$)&  $|R_{\rm F}-R_{\rm M}|$ \\
\hline
3648936 &1.454 &0.013 &1.393 &4.692$\sigma$ \\
4041406 &1.032 &0.008 &1.040 &1.000$\sigma$ \\
4757931 &1.416 &0.013 &1.424 &0.615$\sigma$ \\
4861991 &1.340 &0.021 &1.378 &1.810$\sigma$ \\
5790807 &1.540 &0.020 &1.559 &0.950$\sigma$ \\
6129877 &1.014 &0.008 &0.986 &3.500$\sigma$ \\
6307459 &1.368 &0.013 &1.379 &0.846$\sigma$ \\
8349612 &0.695 &0.000 &1.268 &  --- \\
8745924 &1.264 &0.022 &1.252 &0.545$\sigma$ \\
9204313 &1.133 &0.007 &1.120 &1.857$\sigma$ \\
10644253 &0.923 &0.014 &0.954 &2.214$\sigma$ \\
10454113 &1.253 &0.018 &1.320 &3.722$\sigma$ \\
3221671 &1.277 &0.013 &1.289 &0.923$\sigma$ \\
5295670 &1.466 &0.016 &1.404 &3.875$\sigma$ \\
6285677 &1.338 &0.022 &1.323 &0.682$\sigma$ \\
6306896 &1.402 &0.005 &1.405 &0.600$\sigma$ \\
8561664 &1.073 &0.014 &1.083 &0.714$\sigma$ \\
10454113 &1.253 &0.018 &1.320 &3.722$\sigma$ \\

\hline\hline
\end{tabular}
\end{center}
F=fitted; M=model
\end{table*}

\subsection{Solar data}

\begin{figure}[htbp]
\begin{center}
\includegraphics[width=9cm, trim=28mm 5mm 13mm 15mm]{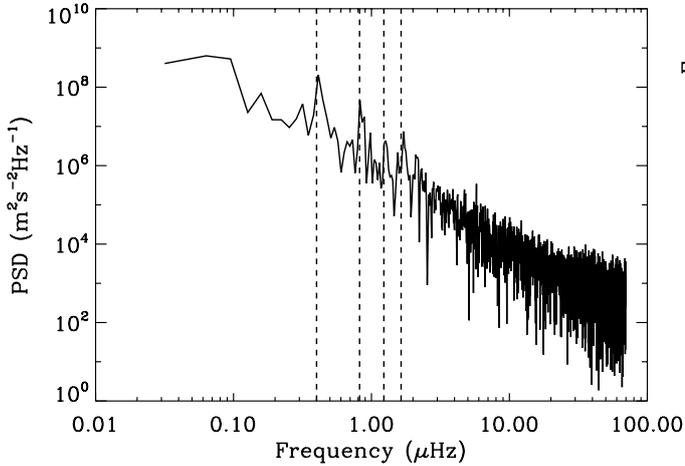}
\caption{Low-frequency part of the PSD obtained with one year of solar data from GOLF instrument as a function of the frequency. The dotted lines highlight the highest peaks below 1~$\mu$Hz that represent the rotation period and the first harmonics.}
\label{psd_sun}
\end{center}
\end{figure}

In this section, we have analysed solar data to test our pipeline on some real data. 
We have estimated the solar surface rotation with the wavelets. We could not use intensity measurements from the VIRGO instrument aboard SoHO as the data available are filtered above 1 and 10 days, which prevents us from measuring the rotation period. Therefore, we have applied the wavelets to velocity data obtained during the first year after the launch of the GOLF (Global Oscillation at Low Frequency) instrument \citep{GarSTC2005}. We have rebinned the data to have one point every 2 hours. We have also applied a filter to remove periodicities above 50 days. First we did a simple fast Fourier transform to calculate the PSD (Fig.~\ref{psd_sun}). We notice four high peaks below 1~$\mu$Hz, at 0.41, 0.8, 1.2, and 1.6~$\mu$Hz, which can be attributed to the rotation period and a high peak at 0.1~$\mu$Hz. However, we cannot disentangle the fundamental period.

\begin{figure}[h!]
\begin{center}
\includegraphics[width=9cm, trim=28mm 5mm 13mm 15mm]{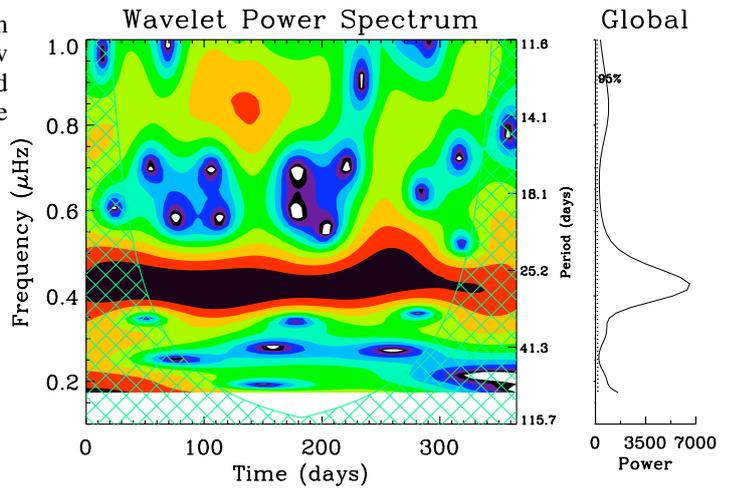}
\caption{Left panel: Wavelet power spectrum as a function of time and frequency of the Morlet wavelet. The period is shown on the right axis of the plot. The green grid represents the cone of influence delimiting the reliable periodicity. Black and red colours denote high-amplitude power whereas green and blue colours highlight low-amplitude power. Right panel: Global wavelet power spectrum as a function of the frequency of the wavelet. The dashed line represents the threshold for a 95\% confidence level. }
\label{fig5b}
\end{center}
\end{figure}

We have then applied the wavelets to these data (Fig.~\ref{fig5b}).  We notice that there is an accumulation of power along time around 0.4-0.5~$\mu$Hz. The COI (green grid) shows that a frequency below $\approx$0.2~$\mu$Hz is not reliable, rejecting thus the peak at 0.1~$\mu$Hz. The Global Wavelet Power Spectrum, which is a collapsogram over time, shows a maximum at 27 days (0.42~$\mu$Hz). The width of the peak is related to differential rotation visible at the surface of the Sun. We can also see a peak around 0.8~$\mu$Hz, but in the WPS there is a power excess only for a short period of time, telling us that this is not the fundamental peak. Thus, this tool is very powerful as it enables us to recover without any ambiguity the real rotation period at the surface of the Sun. This was confirmed using several sets of simulated data as well as solar data obtained with GOLF instrument. 

\begin{figure}[h!]
\begin{center}
\includegraphics[width=9cm, trim=28mm 5mm 13mm 15mm]{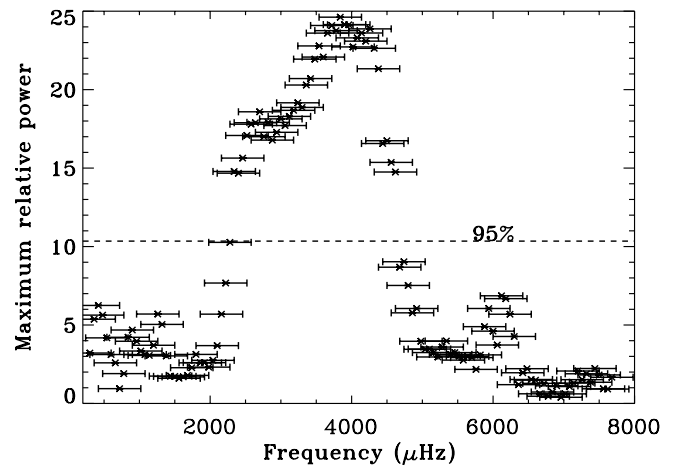}
\caption{Maximum relative power for one month of the VIRGO data of the Sun from package PMRS. Same legend as Fig.~\ref{fig2}. The excess of power of the p modes is in the range 2039 to 4019~$\mu$Hz.}
\label{sun_pmrs}
\end{center}
\end{figure}

\noindent We have used the package PMRS to estimate the mean large separation for the Sun. We have applied it to the VIRGO data, by taking series of 30 days of the blue SpectroPhotoMultiplier, to simulate what we will have with the survey data of Kepler. Using Method 1, we have found $\langle \Delta \nu \rangle$ = 135.46 $\pm$3.3~$\mu$Hz. As the SNR is quite high, it is obtained with a window of 600~$\mu$Hz. There is no double bump. This is also obtained with the highest peak in the PS2. Finally, the p-mode excess power is found between 2039 and 4919~$\mu$Hz (Fig.~\ref{sun_pmrs}). All these values have a confidence level of 95\% and completely agree with the well-known oscillation parameters of the Sun.

\begin{figure}[h!]
\begin{center}
\includegraphics[width=9cm, trim=28mm 5mm 13mm 15mm]{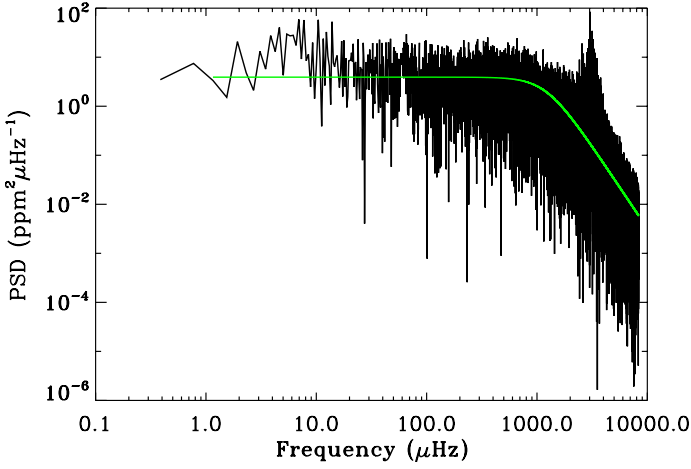}

\caption{Background fitting (green curve) of the solar PSD for one month of data from VIRGO. }
\label{back_sun}
\end{center}
\end{figure}

Fig.~\ref{back_sun} shows the background fit on the PSD. We have subtracted it from the original PSD and we have calculated the maximum amplitude per radial mode with the package BAL0. For the instrumental response of VIRGO, we have used the response function given in \citet{2009A&A...495..979M}. This gives the amplitude envelope plotted on Fig.~\ref{ampl_corot_sun} (black curve). Thus, we obtain $A_{\rm max}$=2.99 $\pm$0.13 ppm and $\nu_{\rm max}$= 3082 $\pm$1.08~$\mu$Hz (Table~4).

Top panel of Fig.~\ref{var_dnu_corot} shows the variation of the large separation with frequency. Our results are compatible with those of \citet{1997SoPh..170....1F}.

The result for the radius of the Sun is given in Tables~5 and 6 and we see that it is fully retrieved within the error bar.

\subsection{Solar-like oscillating stars observed by CoRoT}

We have applied our pipeline to three targets of the CoRoT mission: HD49933 \citep{2008A&A...488..705A, 2009A&A...507L..13B}, HD181906 \citep{2009A&A...506...41G}, and HD181420 \citep{2009A&A...506...51B}. For HD49933, two runs are available: (1) 60 days and (2) 137 days. For HD181906 and HD181420, the data sets are 156 days long.

\begin{figure}[htbp]
\begin{center}
\includegraphics[width=9cm, trim=28mm 5mm 13mm 15mm]{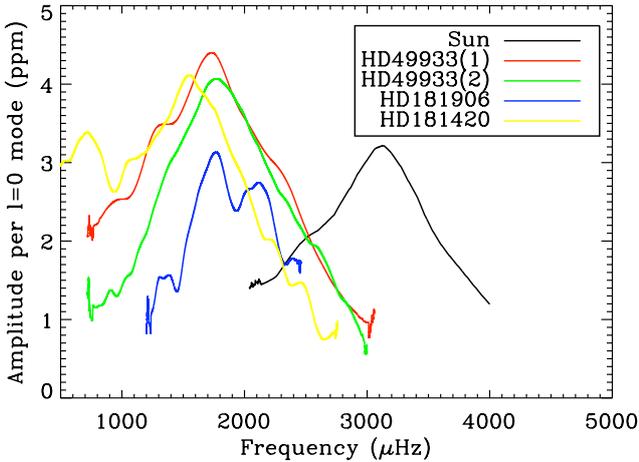}

\caption{Smoothed bolometric amplitude per radial mode (rms) for the Sun, HD49933 (run 1 and 2), HD181906, and HD181420.}
\label{ampl_corot_sun}
\end{center}
\end{figure}

The results of the pipeline are summarised in Tables~4, 5, and 6. The large separations and the p-mode range were obtained with the Method 1 of PMRS (Table~4). They all agree with results already obtained in published papers. Fig.~\ref{ampl_corot_sun} shows the bolometric amplitude in the frequency range where the p-mode bump was found for all these stars.

\begin{table*}[htdp]
\caption{Pipeline results for real data: rotation period, mean large separation, frequency-range of the p modes, position of maximum amplitude, and bolometric maximum amplitude per radial mode.}
\begin{center}
\begin{tabular}{cccccccccc}
\hline
\hline
Star& Period (days)&$\langle \Delta \nu \rangle$ ($\mu$Hz) & $\epsilon$ ($\mu$Hz)& $f_{\rm min}$ ($\mu$Hz) & $f_{\rm max}$ ($\mu$Hz)& $\nu_{\rm max}$ ($\mu$Hz)& $\epsilon$ ($\mu$Hz)& A$_{\rm max}$ (ppm)&  $\epsilon$ (ppm)\\
\hline
Sun & 27.8 & 135.46 & 3.26 & 2040&4920 & 3074.7&1.02 & 2.99& 0.13 \\
HD49933 (1) & 3.2 & 86.14& 1.58& 720& 3060& 1699.88& 1.05&4.08 &0.15\\
HD49933 (2) & 3.3 & 86.33& 1.73& 720& 3000& 1804.34& 0.28&3.90 &0.08\\
HD181906& 2.8 & 86.53& 1.69& 1200& 2460& 1896.9& 1.2& 2.85& 0.13\\
HD181420&  2.6 & 75.35& 1.53& 480& 2760& 1536.26& 0.42 & 3.89&0.1\\
\hline
\hline
\end{tabular}
\end{center}
\label{pipeline_res}
\end{table*}%

\begin{figure}[h]
\begin{center}
\includegraphics[width=9cm]{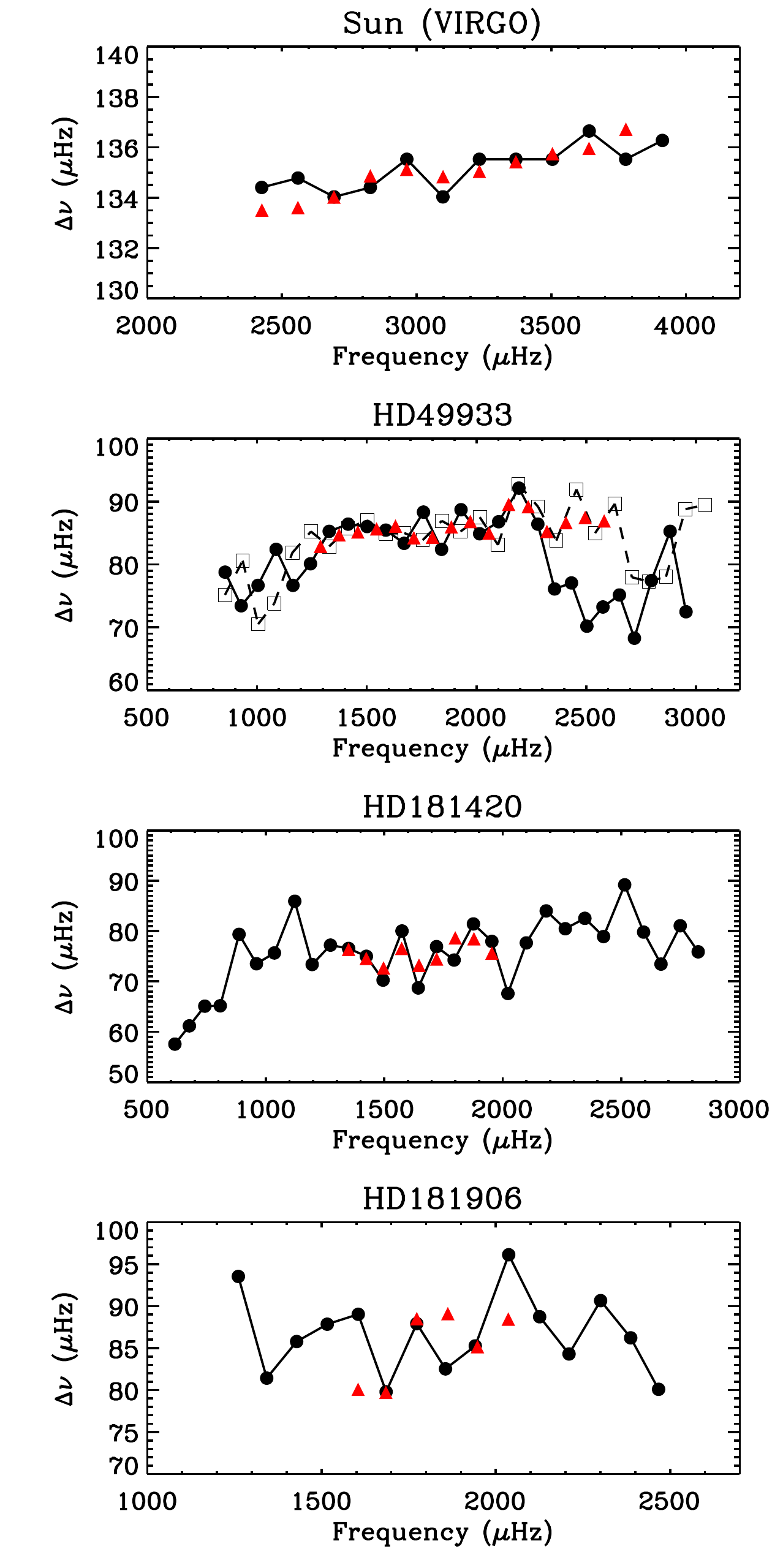}
\caption{Frequency dependence of the large separation obtained as in Sect.~2.5 in the case of the Sun (by using 30 days of VIRGO observations) and three COROT stars: HD49933, HD181420, HD181906 (circles). The results based on published frequencies are also shown (triangles). In the case of HD49933, the squares represent the results obtained from the second run of observations. Only the group of even modes $\ell$=0-2 are represented.}
\label{var_dnu_corot}
\end{center}
\end{figure}

The frequency dependence of the large separation extracted (Sect.~2.5) is illustrated in Fig.~\ref{var_dnu_corot}. The large separations thus measured were compared with published frequency tables   \citep{2009A&A...506...41G,2009A&A...506...51B,2009A&A...507L..13B}. They provide a qualitative good first estimates of their frequency dependence.

With the package FDRM, we have estimated the mass and radius of the stars (Table~\ref{fdrm_real}). Within the error bars, they are compatible with the data we have with spectroscopic observations. So we can have a rough estimation of the mass and radius if we determine $\langle \Delta \nu \rangle$, and the frequency range of the p modes.

\begin{table*}[htdp]
\caption{Estimation of the radius (\emph{R'}) and the mass (\emph{M'}) of real data with the package FDRM compared to spectroscopic observations (R(obs) and M(obs)). }
\begin{center}
\begin{tabular}{ccccccc}
\hline
\hline
Star& \emph{R'} ($R_\odot$)&$\epsilon$ ($R_\odot$)&\emph{R} (obs) ($R_\odot$)& \emph{M'} ($M_\odot$)&$\epsilon$ ($M_\odot$)& \emph{M} (obs) ($M_\odot$)\\
\hline
Sun & 1.003 & 0.1 &1& 1.0167 & 0.2&1 \\
HD49933 (1) & 1.483 &0.15&1.34$\pm$0.016& 1.3279&0.26& 1.2\\
HD49933 (2) & 1.567 &0.15 &1.34$\pm$0.016& 1.5741& 0.35& 1.2 \\
HD181906& 1.581 & 0.16& 1,392 $\pm$0.054& 1.6231&0.32 &1.144 $\pm$0.119 \\
HD181420&  1.7255 &0.17 &1.595 $\pm$0.032& 1.600&0.32& 1.43 $\pm$0.05 \\
\hline
\hline
\end{tabular}
\end{center}
\label{fdrm_real}
\end{table*}%

Using the package RadEx, we have estimated the radius of the CoRoT stars (Table~6). For HD49933, we find a radius of R=1.367 $\pm$0.016 $R_\odot$, while from spectroscopic measurements, R=1.34~$\pm$0.016 $R_\odot$. For HD181906, we obtain R=1.415~$\pm$0.016~$R_\odot$, to be compared to  R=1.392~$\pm$0.054 $R_\odot$. Finally for HD181420, spectroscopic observations have given R=1.595~$\pm$0.032 $R_\odot$ we obtain R=1.623 $\pm$0.026 $R_\odot$. All of these values are in agreement.

\begin{table*}
\begin{center}
\caption{Radius and fitted parameters for real data from stellar
modelling (package RadEx). \label{table:radius_real}}
\begin{tabular}{lcccccccc}
\hline\hline
Star &$R_{\rm F}$ ($R_\odot$)&$\sigma(R)$ ($R_\odot$)&  &  $T_{\rm eff, F}$ (K)&  &
$\langle\Delta\nu\rangle_{\rm F}$ ($\mu$Hz)&
$\langle\Delta\nu\rangle_{\rm O}$ ($\mu$Hz)&
$\epsilon_{\langle\Delta\nu\rangle_{\rm Obs}}$  ($\mu$Hz)\\
\hline
Sun &1.016 &0.023 &      &5879 &     &135.46 &134.67 & 3.26\\
HD49933 &1.398 &0.040 &      &6835 &     & 86.14 & 84.60 & 1.58\\
HD181420 &1.613 &0.032 &      &6596 &     & 75.35 & 76.43 & 1.53\\
HD181906 &1.420 &0.018 &  &6294 &   & 86.53 & 86.12 & 1.69\\
\hline\hline
\end{tabular}
\end{center}
F= fitted; O = Observed
\end{table*}

\section{Discussion and conclusions}

We have described all the different packages of our \emph{A2Z} pipeline to determine the global parameters of stellar oscillations: the mean large separation, the frequency range of the power excess due to the p modes, the background, the maximum bolometric amplitude per radial mode, the guess table of frequencies, and to obtain the radius and the mass of the star.

The first most important step of the pipeline is the estimation of the large separation. We have adopted a 3-step strategy by comparing Method 1 and Method 2 of the package PMRS and by interchanging the results of each of them. We have selected the common results of the step where we obtain the highest number of coincidences with the lowest number of zero confidence level. Applying this methodology to simulated stars has showed that we have obtained from 25 to 40\% of common results with errors from 2 to 11\%. An optional step consists in adding up the common results that are in the 2 best steps. This has led to 29 to 40\% common results with errors ranging from 2.8 to 16.8\%. 


In term of magnitude of the stars, our pipeline seems to be more robust with stars of magnitude below 11. 

It is not yet decided if the last step of our strategy, which adds up the common result from the two best steps, will be used in the future as the number of common results does not increase necessarily and the number of errors increases. It will mainly depend on the the data that we will have with Kepler. 

For the estimation of the mode amplitude, we have managed to estimate 73\% within 3~$\sigma$ of the amplitudes for the Aarhus-asteroFLAG simulations.

Concerning the estimation of the radius of the simulated stars using stellar models, we have succeeded in calculating 70\% of the selected simulated stars. 

Using the data of the Sun and solar-like targets of the CoRoT mission, we have retrieved the global parameters already published and we have also estimated the radius and the mass of the stars with very good accuracy.

By applying the \emph{A2Z} pipeline to three CoRoT targets, we found for HD49933 $\langle\Delta\nu\rangle$=86.2~$\mu$Hz, for HD181906, $\langle\Delta\nu\rangle$=86.53~$\mu$Hz, and for 181420, $\langle\Delta\nu\rangle$=75.35~$\mu$Hz. 

We were able to estimate the maximum amplitude per radial mode for these stars: from 3.9 to 4.08~ppm for HD49933 (depending on the run), 2.85~ppm for HD181906, and 3.89~ppm for HD191420.

We estimated roughly the mass and the radius of these CoRoT targets with the FDRM package, which are, within the large error bars, compatible with the published values from spectroscopic observations.
The radii retrieved from stellar models are in agreement with published results: for HD49933, R=1.367 $\pm$0.016 $R_\odot$, for HD181906, R=1.415~$\pm$0.016~$R_\odot$, and for HD181420, R=1.623 $\pm$0.026 $R_\odot$.

We have to keep in mind that our codes have been modified and tuned to give the best results with the simulated data that we analysed. However, we might have to tune them again with real data from Kepler. So they are not unchangeable and they might evolve in the next months.

 We have already applied this pipeline to one month of Kepler data and obtained some promising results with solar-type stars and  red giants  \citep{BedHub10,ChapApp10,SteBas10}.

\begin{acknowledgements}
This work has been partially supported by: the CNES/GOLF grant at the Service d'Astrophysique (CEA/Saclay) and the grant PNAyA2007-62650 from the Spanish National Research Plan. This work benefited from the support of the International Space Science
Institute (ISSI), through a workshop programme award. It was also partly
supported by the European Helio- and Asteroseismology Network (HELAS), a
major international collaboration funded by the European Commission's
Sixth Framework Programme. WJC acknowledges support from the UK  Science and Technology Facilities Council (STFC).
CoRoT (Convection, Rotation, and planetary Transits) is a mini-satellite developed by the French Space agency CNES in collaboration with the Science Programs of ESA, Austria, Belgium, Brazil, Germany, and Spain \citep{2006cosp...36.3749B}.

\end{acknowledgements}

\bibliographystyle{aa} 
\bibliography{/Users/Savita/Documents/BIBLIO_sav}  

\end{document}